\documentclass[sigconf]{acmart}

\usepackage{amsmath}
\usepackage{booktabs}
\usepackage[T1]{fontenc}
\usepackage{placeins}
\usepackage{bbding}
\usepackage{float}
\usepackage{multirow} 
\usepackage{makecell}
\usepackage{colortbl}
\usepackage{balance}

\AtBeginDocument{%
  }

\copyrightyear{2025}
\acmYear{2025}
\setcopyright{acmlicensed}\acmConference[MM '25]{Proceedings of the 33rd ACM International Conference on Multimedia}{October 27--31, 2025}{Dublin, Ireland}
\acmBooktitle{Proceedings of the 33rd ACM International Conference on Multimedia (MM '25), October 27--31, 2025, Dublin, Ireland}
\acmDOI{10.1145/3746027.3755574}
\acmISBN{979-8-4007-2035-2/2025/10}

\acmSubmissionID{4556} 

\begin{document}

\title{Detect Any Sound:  Open-Vocabulary Sound Event Detection 
with Multi-Modal Queries}

\author{Pengfei Cai}
\email{cqi525@mail.ustc.edu.cn}
\affiliation{
  \institution{University of Science and Technology of China}
  \city{Heifei}
  \state{Anhui}
  \country{China}
}
\author{Yan Song}
\email{songy@ustc.edu.cn}
\affiliation{
  \institution{University of Science and Technology of China}
    \city{Heifei}
    \state{Anhui}
    \country{China}
}

\author{Qing Gu}
\email{qinggu6@mail.ustc.edu.cn}
\affiliation{
  \institution{University of Science and Technology of China}
  \city{Heifei}
  \state{Anhui}
  \country{China}
}

\author{Nan Jiang}
\email{jiang_nan@mail.ustc.edu.cn}
\affiliation{
  \institution{University of Science and Technology of China}
  \city{Heifei}
  \state{Anhui}
  \country{China}
}

\author{Haoyu Song}
\email{2303822@sit.singaporetech.edu.sg}
\affiliation{
  \institution{Singapore Institute of Technology}
  \city{Singapore}
  \country{Singapore}
}

\author{Ian McLoughlin}
\email{Ian.McLoughlin@singaporetech.edu.sg}
\affiliation{
  \institution{Singapore Institute of Technology}
  \city{Singapore}
  \country{Singapore}
}
\renewcommand{\shortauthors}{Pengfei Cai et al.}

\begin{abstract}
Most existing sound event detection~(SED) algorithms operate under a closed-set assumption, restricting their detection capabilities to predefined classes.
While recent efforts have explored language-driven zero-shot SED by exploiting audio-language models, their performance is still far from satisfactory due to the lack of fine-grained alignment and cross-modal feature fusion. 
In this work, we propose the Detect Any Sound Model (DASM), a query-based framework for open-vocabulary SED guided by multi-modal queries. 
DASM formulates SED as a frame-level retrieval task, where audio features are matched against query vectors derived from text or audio prompts.
To support this formulation, DASM introduces a dual-stream decoder that explicitly decouples event recognition and temporal localization: a cross-modality event decoder performs query-feature fusion and determines the presence of sound events at the clip-level, while a context network models temporal dependencies for frame-level localization.
Additionally, an inference-time attention masking strategy is proposed to leverage semantic relations between base and novel classes, substantially enhancing generalization to novel classes.
Experiments on the AudioSet Strong dataset demonstrate that DASM effectively balances localization accuracy with generalization to novel classes, outperforming CLAP-based methods in open-vocabulary setting (+ 7.8 PSDS) and the baseline in the closed-set setting (+ 6.9 PSDS).
Furthermore, in cross-dataset zero-shot evaluation on DESED, DASM achieves a PSDS1 score of 42.2, even exceeding the supervised CRNN baseline. The project
 page is available at \url{https://cai525.github.io/Transformer4SED/demo_page/DASM/}. 


\end{abstract}



\maketitle

\section{Introduction}
Sound event detection (SED) aims to recognize what is happening in an audio signal and when it is happening~\cite{sed-tutorial}.  
As a fundamental task in computer audition, SED has been widely applied to various domains, such as security surveillance~\cite{clavel2005events} and autonomous driving~\cite{CASTORENA}.  
Recently, SED has also gained attention in the development of audio and multimodal large language models~(LLM), as it enables LLMs to perceive environmental audio information~\cite{AudioGPT, Qwen-Audio, an2024funaudiollmvoiceunderstandinggeneration}.  

Existing SED methods~\cite{crnn, nam22_interspeech, li2023ast, ATST} are limited to detecting a predefined set of sound classes, failing to identify novel events unseen during training.  
However, real-world soundscapes consist of thousands of sound classes, whereas SED datasets annotate only a small fraction of these classes.   
In recent years, the audio community has made significant efforts in developing large-scale datasets~\cite{gemmeke2017audio,vgg_sound, audioset-strong} that encompass hundreds of sound classes.
Despite their scale, these datasets often suffer from a severe long-tail distribution, where  rare classes contain only a few minutes of annotated audio, leading to suboptimal  performance due to data scarcity.  
Furthermore, even though these datasets encompass a wide range of sound events, they still cannot guarantee full coverage of all potential events required in applications.  
Consequently, models pre-trained on large-scale datasets typically require fine-tuning on target datasets when applied to new scenarios~\cite{li2023ast, ATST}. 
Constructing scenario-specific datasets and the associated fine-tuning process are both costly and time-consuming, hindering the practicality of SED for real-world applications.

To overcome these limitations, the concept of open-vocabulary learning has been introduced.  
In contrast to traditional closed-set methods, open-vocabulary learning enables models to recognize novel classes beyond the annotated label space.
The core idea is to align audio and language representations through pre-trained audio-language models~\cite{su2025}, allowing models to generalize to novel classes via natural language descriptions.  
Recently, Contrastive audio-language Pretraining (CLAP)~\cite{MS-CLAP, LAION-CLAP} has demonstrated strong open-vocabulary recognition capabilities in the audio classification task by pretraining on large-scale audio-text datasets. 
Some studies have attempted to apply CLAP directly to the SED task~\cite{MGA-CLAP}, but their performance to date is significantly lower than that of closed-set SED models.
This gap can be attributed to two primary factors.  
First, CLAP is trained using clip-level contrastive learning, aligning text and audio at the clip-level without frame-level supervision, which limits its ability to accurately localize sound events.  
Second, the original CLAP framework aligns text and audio only during loss computation, lacking a cross-modal fusion structure to refine representations across modalities.  

In this paper, we introduce Detect Any Sound Model (DASM), a query-based framework for open-vocabulary SED.  
DASM is trained on large-scale SED datasets and combines the precise event localization of closed-set models with the ability to generalize to novel categories.
To achieve these objectives, we formulate open-vocabulary SED as a frame-level retrieval task, where sound events are detected by matching query vectors with frame-wise audio features.  
Specifically, an audio encoder extracts frame-level feature sequences from input audios, while a CLAP based query generation module generates query vectors from event queries.  
Multimodal queries are supported by DASM, allowing queries to be either natural language descriptions (e.g., ``Sound of cats'') or audio clips containing the target event.  
Subsequently, a dedicated decoder architecture is design to match query vectors with audio features.
The decoder in DASM adopts a dual-stream structure to explicitly decouple event recognition and localization: a cross-modality event decoder performs clip-level event recognition via query-feature fusion, while a context network models temporal dependencies for frame-level localization. 
Furthermore, we propose an inference-time attention masking strategy to leverage semantic relations between base and novel classes, substantially aiding generalization to novel events.

To comprehensively evaluate the proposed approach, we construct a benchmark consisting of three subtasks: open-vocabulary detection, closed-set detection, and cross-dataset detection.  
In open-vocabulary experiments on the AudioSet-Strong dataset, DASM achieves a PSDS score of 33.9 on novel classes, surpassing the best CLAP-based approach by 7.8, highlighting the superior capability of our model in open-vocabulary scenarios. 
Under the closed-set setting, DASM attains a PSDS score of 50.9, outperforming closed-set baselines by 6.9 points.  
Furthermore, in cross-dataset evaluation on DESED, DASM achieves a zero-shot PSDS1 score of 42.2, exceeding the supervised CRNN baseline, demonstrating strong generalization across datasets.

\section{Related Work}
\subsection{Sound Event Detection}
\begin{figure}[t]
  \centering
  \includegraphics[width=\linewidth]{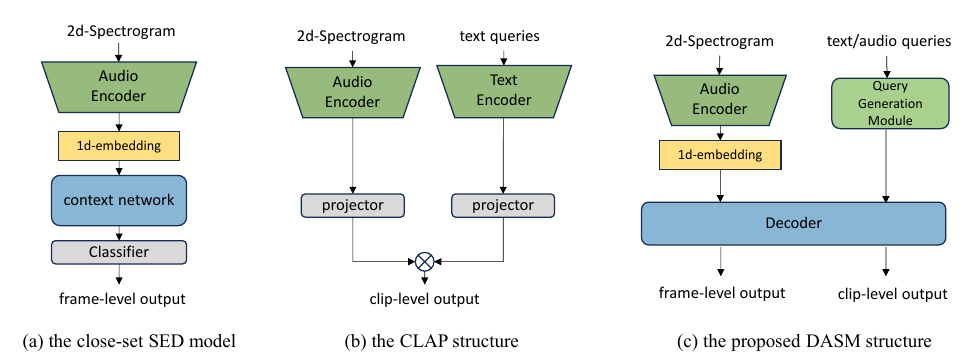}
  \caption{Comparison of three architectures. \textbf{(a)} A classic closed-set SED model, unable to detect novel classes unseen during training.
\textbf{(b)} CLAP architecture, which enables novel class recognition but  is difficult to generate frame-level predictions.
\textbf{(c)} The proposed framework, which detects sound events specified by text or audio queries, enabling novel class detection with frame-level prediction.}
  \label{fig:compare}
\end{figure}

Deep neural network based approaches, represented by CRNNs~\cite{crnn}, have been widely adopted for sound event detection due to their strong audio modeling capabilities.   
CRNN follows the framework illustrated in Fig.~\ref{fig:compare}~(a), where a convolutional neural network (CNN) serves as an encoder to extract acoustic feature sequences along the time dimension, and a recurrent neural network (RNN) based context network to model temporal dependencies of audio features.
The RNN output is passed through a linear classifier followed by a sigmoid function, to yield frame-level event probabilities, denoted as $\textbf{y} \in \mathbb{R}^{T\times C}$, where $T$ represents the number of frames and $C$ is the number of event classes.  
By supporting multi-label classification, CRNNs can effectively handle overlapping sound events, often referred to as the cocktail party problem~\cite{sed-tutorial}.  
Although alternative architectures have been explored beyond the CRNN framework~\cite{ye2021soundeventdetectiontransformer, DiffSED}, these methods have not demonstrated significant performance improvements over CRNN-based architectures and have thus not been widely adopted.

Building on CRNN frameworks, researchers have introduced various enhancements to improve modeling capabilities.
One key challenge in SED is the varying durations and characteristics of different sound events, which CNNs with fixed receptive fields struggle to handle. 
To mitigate this issue, several studies have proposed enhanced CNN architectures~\cite{zhengxu, feature-pyramid-sed, nam22_interspeech} that better accommodate events of varying durations.
In recent years, with the development of Transformer architectures in audio fields~\cite{gong21b_interspeech, pmlr-v202-chen23ag}, works such as~\cite{li2023ast, ATST} have explored using Transformer networks pre-trained on large-scale audio datasets~\cite{gemmeke2017audio} as encoder backbones, to extract generalizable and effective representations. 
Transformer architectures have also been explored as alternatives to RNNs~\cite{Miyazaki2020, wakayama2022cnn, mat-sed, UnifiedAudioEventDetection}, demonstrating superior long-range sequence modeling capabilities.  
However, all of the aforementioned approaches are designed for closed-set scenarios, lacking the flexibility to generalize to unseen categories.  
In this work, we propose a query-based framework that enables open-vocabulary SED with multi-modal queries, paving the way toward universal sound event detection.

\subsection{Open-Vocabulary Learning}
The concept of open-vocabulary learning was first introduced in ~\cite{OVSceneParsing}, requiring models to recognize novel categories beyond the annotated label space.
Unlike zero-shot learning, which involves no exposure to novel classes during training, open-vocabulary learning allows models to utilize language-related annotations, such as image/audio captions, as auxiliary supervision~\cite{ov-overview}.
In recent years, vision-language models (VLMs), exemplified by CLIP~\cite{clip} and ALBEF~\cite{huang2022amae}, have rapidly advanced, bridging the gap between text and images. 
Leveraging the alignment capabilities of VLMs, a variety of open-vocabulary approaches have been proposed, particularly in tasks such as object detection~\cite{OVD, ViLD, OVDETR, GLIP} and semantic segmentation~\cite{LSeg, SegmentAnything}.

\begin{figure*}[t]
  \centering
  \includegraphics[width=0.72\linewidth]{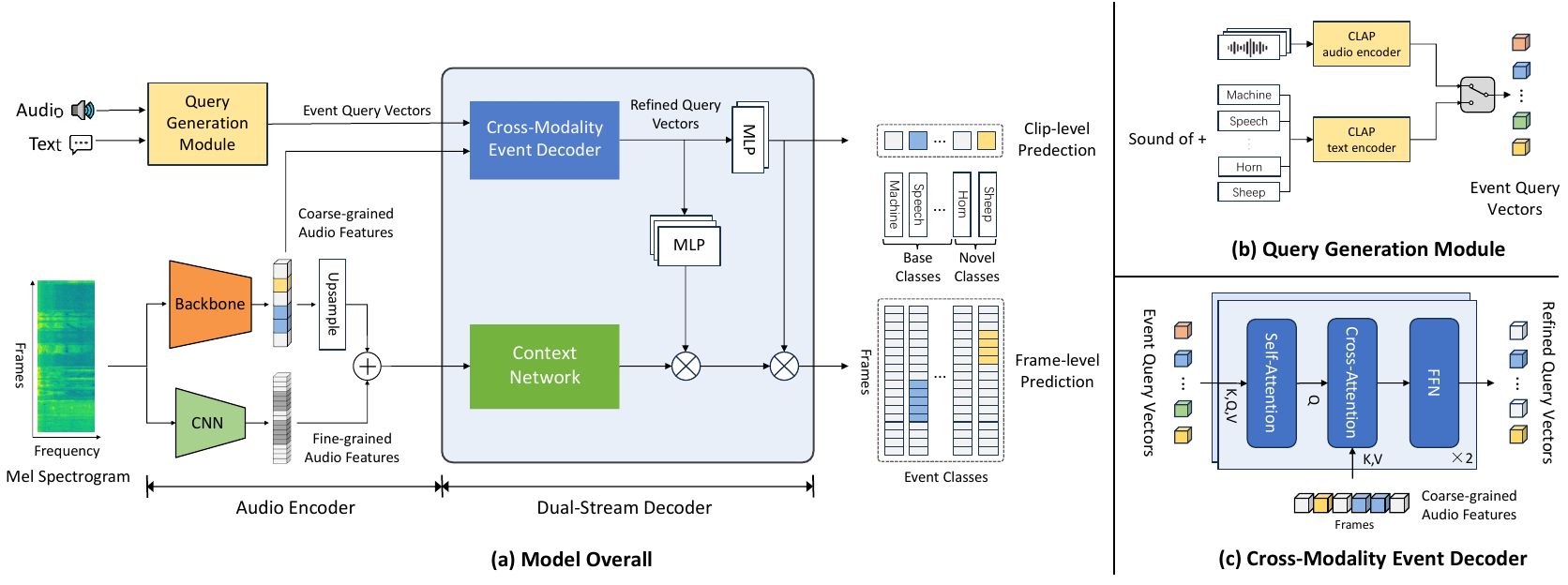}
  \caption{Detect Any Sound Model~(DASM) enables the detection of sound events based on arbitrary text or audio queries. The overall framework, query generation module, and cross-modality event decoder module are presented in (a), (b), and (c), respectively. Note that the channel dimension of features and query vectors is omitted in the figure for simplicity.
  }
  \label{fig:structure}
\end{figure*}

In the audio domain, MS-CLAP~\cite{MS-CLAP} pioneered the pre-training of audio-language models on large-scale audio-text datasets, achieving impressive zero-shot audio recognition performance.  
Subsequent works, including LAION-CLAP~\cite{LAION-CLAP} and WavCaps~\cite{wavCaps}, extended this approach by scaling up training data.  
However, these models primarily focus on clip-level tasks, such as audio classification and retrieval, without adapting to fine-grained tasks like SED.  
To address this issue, Shimada et al.~\cite{Shimada2024} proposed a CLAP-based method for zero-shot sound event localization and detection (SELD), but their approach relies solely on cosine similarity between query and audio features without explicit cross-modal fusion, limiting its expressiveness. 
MGA-CLAP~\cite{MGA-CLAP} introduced a modality-shared codebook to improve local feature alignment and showed improvement for the SED task. 
Nonetheless, it retains the dual-branch architecture and clip-level contrastive training paradigm of CLAP, resulting in a considerable performance gap compared to closed-set SED models.
In contrast, the proposed DASM focuses specifically on the open-vocabulary SED task.
The novel dual-stream decoder enhances both temporal localization accuracy and cross-modal feature fusion, leading to significant performance improvements over prior CLAP-based architectures in open-vocabulary SED.

\section{Methodology}
The core idea of DASM is to treat the open-vocabulary sound event detection task as a frame-level retrieval task.
An overview of the model architecture is shown in Fig. \ref{fig:structure}. 
It consists of an audio encoder that encodes the detected audio into a frame-level feature sequence (Sec. \ref{Audio Encoder}), a query generation module that encodes text or audio queries into event query vectors (Sec. \ref{Query Generation Module}), and a dual-stream decoder that performs query matching between event query vectors and the audio feature sequence (Sec. \ref{Dual-Stream Decoder}).
The input to the model consists of an audio clip to be analysed and event queries, which can either be a textual description of the target event or several audio clips containing the target event.
The dual-stream decoder matches the audio feature sequence with query vectors, determining whether, and  when the events described by the query vectors are active in the detected audio clip.
DASM outputs both clip and frame-level predictions, where the clip-level prediction determines whether the events described by the queries exist in the audio, while the frame-level prediction provides the probability of sound events in each frame.

\subsection{Audio Encoder} \label{Audio Encoder}
The purpose of the audio encoder is to encode the input audio into a latent feature sequence along the temporal dimension.
The audio is pre-processed through short-time Fourier transform (STFT) and Mel filter banks to generate a Mel-spectrogram, which serves as the input to the audio encoder.
Following prior works in SED~\cite{ATST, Cornell2024}, we adopt a dual-branch audio encoder structure, as shown in Fig.~\ref{fig:structure}~(a).
A pre-trained audio spectrogram Transformer model, such as HTS-AT~\cite{htsat}, serves as the backbone, which produces generalizable audio features, denoted as $\textbf{E}_{coarse-grained}$.
However, features from Transformer-based encoder suffer from low temporal resolution, making it challenging to precisely detect event time boundaries.
To address this limitation, a lightweight CNN branch is introduced to generate high-resolution features, denoted as $\textbf{E}_{fine-grained}$, which enhances the accuracy of boundary detection. 
The coarse-grained features $\textbf{E}_{coarse-grained}$ are upsampled via linear interpolation and then combined with $\textbf{E}_{fine-grained}$ to form the final output of the audio encoder, denoted as $\textbf{E}=[\textbf{e}_1, \textbf{e}_2, \ldots, \textbf{e}_{N}] \in \mathbb{R}^{T \times D}$:
\begin{equation}
\textbf{E} = \textbf{E}_{fine-grained} + {Upsample}(\textbf{E}_{coarse-grained}) 
\end{equation}
Here, $T$ represents the number of frames, $D$ represents the feature dimension, and $\textbf{e}_i$~($i \in[1, T]$) is the audio feature embedding for the $i$-th time frame.

\subsection{Query Generation Module} \label{Query Generation Module}
The query generation module is responsible for generating query vectors corresponding to the events of interest. 
The core component of this module is the pre-trained CLAP model, as shown in Fig.~\ref{fig:structure}~(b).
When the number of queried events is $N$, the query generation module outputs $N$ event query vectors, denoted as $\textbf{Q} = [\textbf{q}_1, \textbf{q}_2, \ldots, \textbf{q}_N] \in \mathbb{R}^{N \times D}$. 
Since CLAP can align text and audio modalities into the same feature space, queried events can be provided in either textual or audio form.
When the event is described with text form, we use the prompt template ``sound of \{class\}'' to convert the event name into a sentence, which is then fed into CLAP's text encoder to obtain the corresponding query vector.
When the event is given in audio form, i.e., when several audio clips containing the target event are available as prompts, we first extract the segments that contain the target event. 
These audio segments are then input into CLAP’s audio encoder. 
The output features are pooled along the time dimension using average pooling to obtain the query vector for the audio event.
In practice, we find that a few minutes of audio are sufficient to construct representative query vectors.

During training, the query generation process is performed offline. 
When the DASM model is trained to support both textual and audio queries, the query generation module randomly selects between the two modalities of query vectors with a probability of 0.5.
During inference, queries for novel classes are concatenated after the base query sequence, thus enabling the model to handle open-vocabulary tasks.

\subsection{Dual-Stream Decoder} \label{Dual-Stream Decoder}
Through the audio encoder and the query generation module, we obtain the audio feature sequence $\textbf{E}$ and the query vectors $\textbf{Q} = [\textbf{q}_1, \textbf{q}_2, \ldots, \textbf{q}_N]$.
Previous works, as shown in Fig.~\ref{fig:compare}~(b), have obtained predictions by computing the cosine similarity between the query vectors and the audio features of different frames, where the query vector serves as the classifier in closed-set SED models. This approach only involves interaction between the two branches during loss computation and lacks a cross-modal feature fusion structure.
In contrast, we introduce a decoder structure for tight feature fusion and fine-grained feature alignment.
The decoder adopts a dual-stream structure, as shown in Fig.~\ref{fig:structure}~(a), where the cross-modality event decoder performs query matching with the audio feature sequence for clip-level event recognition, while the context network focuses on modeling the temporal dependencies of the features to improve the accuracy of event boundary localization.
The results of both branches are combined to produce the frame-level event predictions.

\subsubsection{Cross-Modality Event Decoder}
The cross-modality event decoder, as shown in Fig.~\ref{fig:structure}~(c), is stacked by two standard Transformer decoder blocks~\cite{vaswani2017attention}.
Each decoder block contains a self-attention layer, a cross-attention layer, and a feedforward network (FFN) layer, with residual connections and layer normalization applied to all three layers. 
Similarly to \cite{DETR}, the computation of different queries is executed in parallel, without requiring causal attention masking.
In the self-attention layer, the model computes attention scores between queries, enabling event query vectors to focus on other events that are semantically related. 
This helps the model capture hierarchical relationships and semantic correlations between sound events, as detailed in \ref{training and inference}.
In the cross-attention layer, the query vectors act as the queries, while the coarse-grained audio features $\textbf{E}_{coarse-grained}$ from the backbone serve as the keys and values.
Through the cross-attention mechanism, frames containing the target event get higher attention scores, allowing their features to be selectively aggregated into the query vectors. 

The cross-modality event decoder outputs refined query vectors, denoted as $\textbf{Q}_{refined} = [\textbf{q}_{refined,1}, \textbf{q}_{refined,2}, \ldots,  
 \textbf{q}_{refined,N}] \in \mathbb{R}^{N \times D}$.
The refined query vectors benefit from feature fusion with the audio features and between other 
event queries, thus serve as better classifiers than original query vectors.
Subsequently, the refined query vectors are passed through a two-layer MLP classifier $MLP_{cl}$, for binary classification, yielding the probability that an event occurs in the audio clip:
\begin{equation}
y^{cl}_{i} = \mathbb{P}(Y^{cl}_{i} = 1 |\textbf{x}) = sigmoid(\mathrm{MLP}_{cl}(\textbf{q}_{refined,i}))
\end{equation}
Here, the random variable $Y^{cl}_{i} \in \{0, 1\}$ represents whether the $i$-th sound event occurs in the audio clip, with 1 for occurs and 0 for not, and $\textbf{x}$ represents the input audio to be detected.
In parallel, the refined query vectors are also mapped through a three-layer MLP ($\mathrm{MLP}_{map}$) to generate classifier vectors, denoted as $\textbf{c}_{i} \in \mathbb{R}^{D}$:
\begin{equation}
 \textbf{c}_{i} = \mathrm{MLP}_{map}(\textbf{q}_{refined,i})
\label{eq:cls}
\end{equation}
The classifier vectors will be used in the next section for frame-level sound event detection.

\subsubsection{Context Network}
The context network processes the audio feature sequence $\textbf{E} = [\textbf{e}_1, \textbf{e}_2, \ldots, \textbf{e}_N]$ to model temporal dependencies across the feature sequence.
We employ two layers of Conformer~\cite{gulati20_interspeech} blocks for this purpose.
The output of the context network maintains the same shape as its input and is denoted as $\textbf{Z} = [\textbf{z}_1, \textbf{z}_2, \ldots, \textbf{z}_N] \in \mathbb{R}^{T \times D}$.

The frame-level prediction is formulated as:
\begin{align}
    y^{fr}_{i,t} &= \mathbb{P}(Y^{fr}_{i,t} = 1|\textbf{x}) \notag \\
    &= \mathbb{P}(Y^{fr}_{i,t} = 1 | Y^{cl}_{i} = 1,\textbf{x}) \mathbb{P}(Y^{cl}_{i} = 1|\textbf{x})  \notag \\
    &\quad + \mathbb{P}(Y^{fr}_{i,t} = 1 | Y^{cl}_{i} = 0,\textbf{x}) \mathbb{P}(Y^{cl}_{i} = 0|\textbf{x})
    \label{eq:y_frame_origin}
\end{align}
where the random variable $Y^{fr}_{i,t} \in \{0, 1\}$ represents whether the $i$-th sound event occurs at the $t$-th frame in the fine-grained prediction.
Notably, if the event does not occur in the clip, its probability at any frame is zero:
\begin{equation}
\mathbb{P}(Y^{fr}_{i,t} = 1 | Y^{cl}_{i} = 0, \textbf{x}) = 0    
\end{equation}
Thus in Eq.\ref{eq:y_frame_origin}, only $\mathbb{P}(Y^{fr}_{i,t} = 1 | Y^{cl}_{i} = 1, \textbf{x})$ needs to be estimated to derive the frame-level prediction.
We utilize the classifier vector $\textbf{c}_{i}$ obtained from Eq.~\ref{eq:cls} to predict this conditional probability as follows:
\begin{equation}
    \mathbb{P}(Y^{fr}_{i,t} = 1 | Y^{cl}_{i} = 1, \textbf{x}) = sigmoid(\textbf{z}_{t}^{T}\textbf{c}_{i})
    \label{eq:condition}
\end{equation}
Substituting Eq.\ref{eq:condition} into Eq.\ref{eq:y_frame_origin}, the final expression for frame-level probability is:
\begin{equation}
y^{fr}_{i,t} = sigmoid(\textbf{z}_t^{T}\textbf{c}_{i})\cdot y^{cl}_{i}    
\label{eq:y_frame_final}
\end{equation}

Eq.~\ref{eq:y_frame_final} derives the frame-level event prediction by factorizing it into two multiplicative terms: a classifier-based prediction that models the conditional probability of the event occurring at a specific frame, and the clip-level prediction, acting as a prior to incorporate global event occurrence probability.
This factorized computation is a key design that decouples recognizing what event is present from when it occurs, improving both precision and recall for the SED task, as demonstrated in subsequent experiments. 


\begin{figure}[t]
  \centering
  \includegraphics[width=0.7\linewidth]{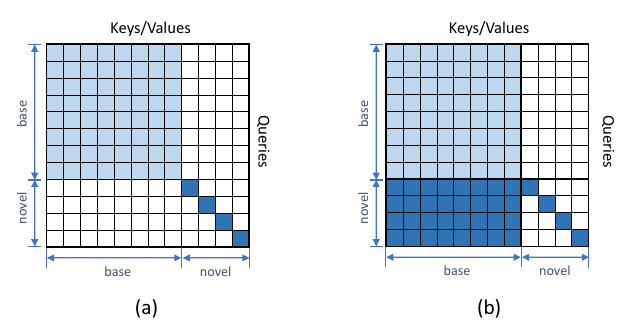}
  \caption{Two attention masking strategies in self-attention layers of the  event decoder during inference.
(a) Novel class query vectors cannot attend to base class query vectors.
(b) Base class query vectors remain visible to novel class query vectors.}
  \label{fig:mask}
\end{figure}

\subsection{Training and Inference} \label{training and inference}
During training, the loss function for an input audio sample $\textbf{x}$ is  formulated  as the sum of the frame-level loss $L_{fr}$ and the clip-level loss $L_{cl}$:
\begin{equation}
    L = \sum_{i=1}^{N}\sum_{t=1}^{T} L_{fr}(y^{fr}_{i,t}, \hat{y}^{fr}_{i,t}) +\alpha\sum_{i=1}^{N} L_{cl}(y^{cl}_{i}, \hat{y}^{cl}_{i})
    \label{equ:loss}
\end{equation}
where $\hat{y}^{fr}_{i,t}$ denotes the frame-level label for the $t$-th frame of the $i$-th event, and $\hat{y}^{cl}_{i}$ represents the clip-level label of the $i$-th event in  $\textbf{x}$.
The weighting coefficient $\alpha$ is set to 0.5 in our experiments empirically. 
Both loss functions adopt the asymmetric focal loss~\cite{ridnik2021asymmetric,focal_sed} to mitigate the class imbalance present in SED datasets.

In open-vocabulary scenarios, the model needs to detect novel classes that were never seen during training.
The primary distinction between training and inference in DASM lies in the self-attention mechanism within the cross-modality event decoder.
During training, self-attention operates without masking, allowing base queries to attend to each other freely. 
During inference, novel queries are appended to the query sequence, and attention masking is applied in the self-attention layers.
Attention masking prevents base classes from attending to novel classes, thereby avoiding potential interference from novel classes that could degrade base class detection.
For novel classes, we explore two masking strategies: base classes remain invisible to novel classes, as illustrated in Fig.~\ref{fig:mask}~(a), and base classes remain visible to novel classes, as illustrated in Fig.~\ref{fig:mask}~(b).
Experimental results demonstrate that allowing base classes to remain visible to novel classes significantly improves open-vocabulary detection  performance, compared to restricting their visibility.
Maintaining base class visibility helps the model exploit the semantic similarity and hierarchical structure between novel and base classes.
For instance, when identifying a novel class such as 'Electric Rotor Drone', the model has never encountered this exact term during training. 
However, within the ontology of sound events, it may share hierarchical relationships with existing base classes, such as its hypernym 'Mechanisms' or a semantically related concept like 'Mechanical Fan.'
The self-attention mechanism facilitates the association between novel classes and relevant base classes, thereby improving performance for unseen sound events.

\section{EXPERIMENTAL SETUP}
To comprehensively evaluate the performance of the proposed DASM, we construct a benchmark comprising three subtasks: open-vocabulary sound event detection, closed-set sound event detection, and cross-dataset detection.
In this section, we first introduce datasets and evaluation metrics, followed by model configuration and training settings.

\subsection{Datasets}
We evaluate our model on AudioSet Strong and DESED, two benchmark datasets in sound event detection. 
In our experiments, we use AudioSet Strong for open-vocabulary and closed-set evaluations, and employ DESED to evaluate the model’s cross-dataset transfer capability.

\paragraph{\textbf{AudioSet Strong}}
AudioSet Strong~\cite{audioset-strong} is a large-scale SED dataset with temporally strong
labels.
In our experiments, we apply the label augmentation using AudioSet-Ontology~\cite{gemmeke2017audio} to embed label hierarchical structure of sound events. 
Specifically, for each annotated sound event, we include all its hypernyms (i.e., parent nodes) in the ontology as additional labels, and remove samples containing events not included in the ontology.
After label augmentation, the dataset contains 407 sound event classes, with 93,927 audio clips in the training set and 14,203 clips in the evaluation set.

The event duration distribution in AudioSet Strong exhibits a typical long-tail pattern.
In the experiments, we classify sound events into common and rare classes based on their total duration in the training set, using a threshold of six minutes, where 99 events with a duration of less than six minutes are designated as rare, while the remaining events are categorized as common.
For open-vocabulary experiments, we follow the partial-label evaluation strategy~\cite{bansal2018zero, ViLD}, treating common classes as base classes while keeping rare classes as novel classes. 
The model is trained only on common classes with all rare class labels removed to ensure zero exposure, while evaluation is conducted on the full label set including both common and rare classes.
This setting is denoted as \textbf{AS-partial}.
For closed-set evaluations, all labels are visible during training, which we denote as \textbf{AS-full}.

\paragraph{\textbf{DESED}}
DESED is a dataset specially designed for detecting sound events in domestic environments, composed of 10 event classes. 
The training set is heterogeneous, containing 1,578 weakly labeled clips, 3,470 strongly labeled clips, 10,000 synthetic strongly labeled clips, and 14,412 unlabeled in-domain clips.
The validation set consists of 1,168 strongly labeled clips.
We use DESED to evaluate the model's cross-dataset  transfer ability.

\begin{table*}[h]
    \centering
    \caption{Experiments on AudioSet-Strong.
    AS-partial denotes the open-vocabulary setting using a partial-label strategy, where rare class labels are removed during training and PSDS$_r$ is used as the primary metric.
    AS-full refers to the closed-set setting with all labels visible, evaluated using the overall PSDS.
    Models with query type `None' are  closed-set models and models marked with * are evaluated in a zero-shot setting without training on AudioSet Strong.
}
    
    \label{tab:as_experiment}
    \setlength{\tabcolsep}{10pt} 
    \small
    \begin{tabular}{l c c|>{\color{gray}}c >{\color{gray}}c c|c >{\color{gray}}c >{\color{gray}}c}
        \Xhline{1.25pt}
        \multirow{2}{*}{\textbf{Method}} & \multirow{2}{*}{\textbf{Backbone}} & \multirow{2}{*}{\textbf{Query}} & \multicolumn{3}{c|}{\textbf{AS-partial}} & \multicolumn{3}{c}{\textbf{AS-full}} \\
        & & & \textbf{PSDS} & \textbf{PSDS$_c$} & \textbf{PSDS$_r$} & \textbf{PSDS} & \textbf{PSDS$_c$} & \textbf{PSDS$_r$} \\
        \Xhline{0.8pt}
        HTS-SED~\cite{ATST, Cornell2024} & HTS-AT & None & - & - & - & 44.0 & 47.3 & 31.8 \\
        LAION-CLAP~\cite{LAION-CLAP} & HTS-AT & Text & 36.0 & 41.4 & 16.1 & 38.1 & 41.4 & 25.7 \\
        MGA-CLAP*~\cite{MGA-CLAP} & HTS-AT & Text & 35.3 & 37.8 & 26.1 & 35.3 & 37.8 & 26.1 \\
        MGA-CLAP~\cite{MGA-CLAP} & HTS-AT & Text & 39.7 & 44.2 & 22.9 & 40.9 & 44.2 & 28.8 \\
        \hline
        DASM & HTS-AT & Text & 48.8 & \textbf{53.9} & 30.1 & 49.8 & 52.9 & \textbf{38.4} \\
        DASM& HTS-AT& Audio & 49.0 & 53.1 & \textbf{33.9} & 49.9 & 53.2 & 37.8 \\
        DASM& HTS-AT& Text/Audio & 48.4 & 52.9 & 31.4 & 49.9 & 53.1 & 37.9 \\
        \hline
        DASM& PaSST  & Text & 47.2 & 53.7 & 23.3 & 50.0 & 53.7 & 36.5 \\
        DASM& PaSST& Audio & \textbf{49.2} & 53.6 & 32.7 &  \textbf{50.9} & \textbf{54.4} & 38.3 \\
        DASM& PaSST& Text/Audio & 47.5 & 53.5 & 25.1 & 50.6 & 54.2 & 37.2 \\
        \Xhline{1.25pt}
    \end{tabular}
\end{table*}

\subsection{Evaluation Metrics}
We evaluate sound event detection performance using the polyphonic sound detection score (PSDS)~\cite{psds}, a widely adopted metric in recent SED research.
Specifically for DESED, we use the $\mathrm{PSDS_1}$ metrics~\cite{Cornell2024}, which emphasizes event-wise 
temporal localization accuracy.
When applied to AudioSet Strong, the class variance penalty of $\mathrm{PSDS_1}$ is omitted following previous works~\cite{schmid2024, morocutti2025exploring}, since it was designed for more balanced datasets.
For completeness, we report the overall PSDS score of AudioSet Strong, as well as separate scores for common~(PSDS$_c$) and rare~(PSDS$_r$) classes.
Under the AS-partial setting, PSDS$_r$ is adopted as the primary metric to evaluate the model’s open-vocabulary detection performance on rare classes. For the AS-full setting, the overall PSDS score is used to assess the model’s closed-set detection capability on all classes.

\subsection{Model Configuration}
\paragraph{\textbf{Query Generation Module}} 
We train three DASM variants based on different query modalities: text-only, audio-only, and a multi-modal version which supports text or audio queries.
For audio queries, segments containing the target events in the training set are used as prompt examples.
A pre-trained MGA-CLAP~\cite{MGA-CLAP} model is used as the query generation module, where HTS-AT serves as the audio encoder~\cite{htsat} and BERT~\cite{DevlinCLT19} as the text encoder.
This module is used for offline query vector extraction and remains frozen during training.

\paragraph{\textbf{Audio Encoder}} 
We use two Transformer-based pre-trained audio models as the backbone: HTS-AT (31M), which features a Swin-Transformer-like hierarchical structure~\cite{swin}, and PaSST (86M), using a ViT-based architecture~\cite{dosovitskiy2021an}.
HTS-AT is initialized using weights from the MGA-CLAP audio encoder, while PaSST is initialized with weights pre-trained in AudioSet.
For PaSST, we additionally apply an attention-based pooling layer to compress the frequency dimension of output features.
The lightweight CNN branch consists of 10 CNN blocks with 1.7M parameters.
The coarse-grained backbone features are merged with fine-grained CNN features, resulting in a final temporal resolution of 50 frames per second.

\paragraph{\textbf{Decoder}} 
The decoder is composed entirely of Transformer blocks, each with an input dimension of 384, 12 attention heads, and a FFN expansion ratio of 1. 
It contains 7.9M parameters in total, considerably lighter than the backbone.

\subsection{Training Settings}
The input audio is sampled at 32 kHz.
Mixup~\cite{zhang2018mixup}, time shift, and SpecAugment~\cite{park19e_interspeech} are used for data augmentation during training.
For AudioSet Strong, the model is trained for 40K steps with a batch size of 32.
During the first 20K steps, the backbone is frozen, and only the CNN encoder and decoder are trained.
The learning rate is set to $1.5\times 10^{-5}$ for the backbone and $2\times 10^{-4}$ for other components.
Optimization is performed using AdamW~\cite{LoshchilovH19} with a weight decay of $1\times 10^{-4}$.
To mitigate the class imbalance in AudioSet Strong, we apply audio resampling based on event occurrence frequency, following \cite{koutini2021efficient}.
For fine-tuning on DESED, we adopt the mean-teacher~\cite{tarvainen2017mean} semi-supervised  strategy, following~\cite{li23n_interspeech}.
During inference, median filter with a window length of 5 frames is used for post-processing to frame-level prediction.

\section{RESULTS}
\subsection{Open-Vocabulary Sound Event Detection}
We compare DASM against CLAP-based approaches to evaluate the performance of  open-vocabulary sound event detection. Specifically, we compare with two CLAP variants: LAION-CLAP~\cite{LAION-CLAP} and MGA-CLAP~\cite{MGA-CLAP}. 
To adapt CLAP for the sound event detection task, we  compute the similarity between text embeddings and the frame-level features. For LAION-CLAP, which inherently supports only clip-level tasks, we perform frame-level fine-tuning on AudioSet Strong using the same training configuration as DASM. 
For MGA-CLAP,  since it natively supports the sound event detection task, we report both its zero-shot and fine-tuning performance.

The results are presented in the AS-partial column of Tab.~\ref{tab:as_experiment}. When using HTS-AT as the backbone, DASM with text queries achieves a $\mathrm{PSDS_r}$ score of 30.1, outperforming LAION-CLAP by 14.0 and MGA-CLAP by 4.0, demonstrating its superior open-vocabulary detection capability. 
Using audio queries further boosts the $\mathrm{PSDS_r}$ score to 33.9, an increase of 3.8 over text queries.
This is likely due to the fact that audio queries and the input audio share the same modality, thus without cross-modal alignment issues. The mixed-modality DASM, which supports text or audio queries, achieves a $\mathrm{PSDS_r}$ score of 31.4, falling between the performance of the text-only and audio-only models.
Generally, the performance across all query modalities remains relatively close.
Notably, despite the model never encountering novel classes during training, DASM achieves even higher performance than the closed-set baseline in the AS-full setting (see Section~\ref{closed-set}). This further underscores DASM’s strong generalization ability for unseen events.

Additionally, we investigate the effect with a larger backbone by replacing HTS-AT with PaSST. However, when using PaSST, the $\mathrm{PSDS_r}$ score for text queries drops to 23.3, which is 6.8 lower than that of the HTS-AT-based model. A possible explanation for this phenomenon is that PaSST is initialized with weights pre-trained on AudioSet for audio classification, lacking explicit alignment between audio and text modalities. This also explains why the performance of DASM with audio queries significantly surpasses that with text queries when using PaSST, as modality misalignment is not an issue for audio queries.

\subsection{Closed-Set Sound Event Detection} \label{closed-set}
In this section, we evaluate the closed-set sound event detection performance of DASM by training the model on all classes in the AudioSet Strong dataset.
In addition to CLAP-based approaches, we compare DASM with traditional closed-set models with structures illustrated in Fig.~\ref{fig:compare}~(a). 
Following previous work~\cite{ATST, Cornell2024}, we design a HTS-AT based closed-set baseline, denoted as HTS-SED, which adopts an identical audio encoder and context network with the HTS-AT based DASM but employs a linear classifier on the top layer, making it strictly a closed-set model.

The closed-set SED experimental results are presented in the AS-full column of Table~\ref{tab:as_experiment}. The closed-set baseline HTS-SED achieves a PSDS of 44.0, which aligns well with results reported in previous works~\cite{schmid2024}. 
This score is significantly higher than those obtained by CLAP-based models, highlighting the limitations of CLAP architectures in sound event detection.
Meanwhile, DASM with an HTS-AT backbone achieves a PSDS of 49.9, surpassing HTS-SED by 5.9, indicating that DASM is also effective in closed-set scenarios.
When using PaSST as a larger backbone, the PSDS further improves to 50.9, demonstrating the effectiveness of model scaling.

\begin{table}[t]
    \centering
    \caption{Experiments on DESED. Models with `None' as the query type are closed-set models.
}
    \small
    \label{tab:psds1_results}
    \begin{tabular}{l c c | c c}
        \Xhline{1.25pt}
        \multirow{2}{*}{\textbf{Method}} & \multirow{2}{*}{\textbf{Backbone}} & \multirow{2}{*}{\textbf{Query}} & \multicolumn{2}{c}{\textbf{PSDS1}} \\
        & & & {\footnotesize \textbf{zero-shot}} & {\footnotesize \textbf{finetune}} \\
        \Xhline{0.8pt}
        {DCASE Baseline~\cite{Cornell2024}} & CNN & None & - & 36.4 \\
        {DCASE Baseline~\cite{Cornell2024}} & BEATs & None & - & 50.0 \\
        PASST-SED~\cite{li23n_interspeech} & PaSST & None & - & 55.5 \\
        ATST-SED~\cite{ATST} & ATST & None & - & 58.3 \\
        LAION-CLAP~\cite{LAION-CLAP} & HTSAT & Text & 22.8 &  44.6\\
        MGA-CLAP~\cite{MGA-CLAP} & HTS-AT & Text & 24.6 & 45.5 \\
        \hline
        DASM & HTS-AT & Text & \textbf{42.2} & 58.6 \\
        DASM & HTS-AT & Audio & 37.8 & \textbf{58.8} \\
        DASM & HTS-AT & Text/Audio & 39.6 & 58.6 \\
        \hline
        DASM & PaSST & Text & 40.7 & 57.1 \\
        DASM & PaSST & Audio & 38.5 & 57.5 \\
        DASM & PaSST & Text/Audio & 39.1 & 56.4 \\
        \Xhline{1.25pt}
    \end{tabular}
\end{table}

\begin{table*}[t]
    \centering
    \begin{minipage}{0.64\textwidth}  
        \centering
        \footnotesize
        \caption{Ablation study on the decoder. All experiments use HTS-AT as the backbone with text-based queries.}
        \label{tab:ablation_text_query}
        \renewcommand\arraystretch{1.2}
        \begin{tabular}{cc|ccc|ccc|c}
            \Xhline{1.25pt}
            \multirow{2}{*}{\textbf{\#ID}} & \multirow{2}{*}{\textbf{Model}} & \multicolumn{3}{c|}{\textbf{AS-partial}} & \multicolumn{3}{c|}{\textbf{AS-full}} & \textbf{DESED} \\
            & & \textbf{PSDS} & \textbf{PSDS$_c$} & \textbf{PSDS$_r$} & \textbf{PSDS} & \textbf{PSDS$_c$} & \textbf{PSDS$_r$} & \textbf{\scriptsize PSDS1(zero-shot)} \\
            \Xhline{0.8pt}
            \rowcolor{gray!10}
            0 & DASM  & \textbf{48.8} & \textbf{53.9} & \textbf{30.1} & \textbf{49.8} & \textbf{52.9} & \textbf{38.4}& \textbf{42.2}\\
            1 & w/o event decoder &39.9 & 44.6& 23.3& 41.0&43.5 & 31.9& 21.8\\
            2 & w/o context network  & 29.3 & 33.4& 13.9& 29.3& 33.3&14.7 & 10.5\\
            3 & w/o clip loss  &43.5 &48.4 &26.0 & 44.9& 48.1& 33.2& 41.6\\
            4 & w/o clip-level prior &44.3 &49.1 &26.8 &45.4 &48.5 &33.7 &36.7 \\
            \Xhline{1.25pt}
        \end{tabular}
    \end{minipage}
    \hfill
    \begin{minipage}{0.32\textwidth}  
        \centering
        \footnotesize
        \caption{Effect of base-class visibility on novel classes detection during inference. The table reports PSDS$_r$ scores under the AS-partial setting, using HTS-AT as the backbone.}
        \label{tab:inference}
        \begin{tabular}{c |c c c}
            \Xhline{1.25pt}
            \multirow{2}{*}{\textbf{\makecell{Visibility of \\base classes}}} & \multicolumn{3}{c}{\textbf{Query type}} \\
            & \textbf{Text} & \textbf{Audio} & \textbf{Text/Audio} \\
            \Xhline{0.8pt}
            \rowcolor{gray!10}
            $\checkmark$ & \textbf{30.1}& \textbf{33.9}& \textbf{31.4}\\
            $\times$& 4.6& 1.0& 7.9\\
            \Xhline{1.25pt}
        \end{tabular}
    \end{minipage}
\end{table*}

\subsection{Cross-Dataset Detection} \label{Cross-dataset}

In this experiment, we assess the cross-dataset generalization of DASM by evaluating models pre-trained on AudioSet Strong directly on the DESED dataset in a zero-shot manner.
In addition to CLAP-based methods, we compare our model's performance with the baseline model used in DCASE 2023 and 2024 challenges~\cite{Cornell2024} and other advanced closed-set models on DESED.
To provide a more comprehensive comparison with closed-set studies, we also report fine-tuned results of CLAP and DASM models on DESED.

As shown in Tab.~\ref{tab:psds1_results}, DASM achieves an impressive zero-shot $\mathrm{PSDS_1}$ score of 42.2, even surpassing the supervised CNN-based baseline by 5.8, demonstrating its strong generalization ability.
In contrast, CLAP-based models, including LAION-CLAP and MGA-CLAP, achieve only 22.8 and 24.6 in $\mathrm{PSDS_1}$, respectively, falling significantly behind the baseline, despite being trained on AudioSet Strong with frame-level supervision.
After fine-tuning on the DESED dataset, DASM further improves its $\mathrm{PSDS_1}$ score to 58.8, outperforming the closed-set models included in our comparison. This further shows the effectiveness of the proposed architecture.

\begin{figure}[t]
  \centering
  \includegraphics[width=0.55\linewidth]{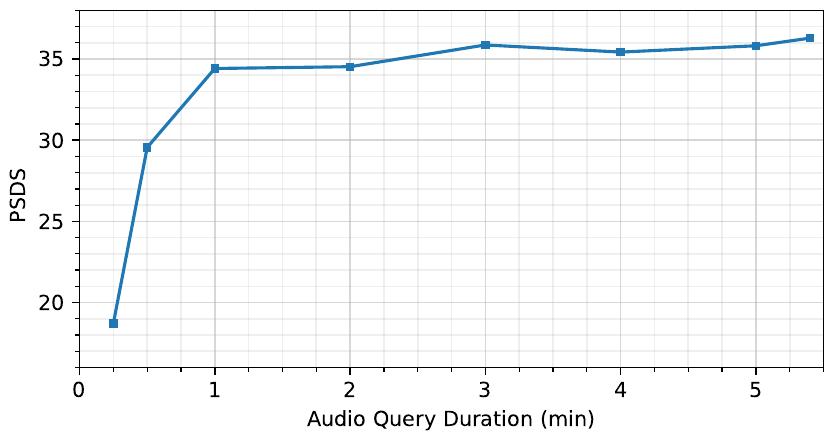}
  \caption{Impact of audio query duration. 
  We evaluate DASM on novel classes with audio query durations ranging from 5 to 6 minutes and progressively reduce the query duration.  
}
  \label{fig:duration}
\end{figure}

\begin{figure*}[t]
  \centering
  \includegraphics[width=0.8\linewidth]{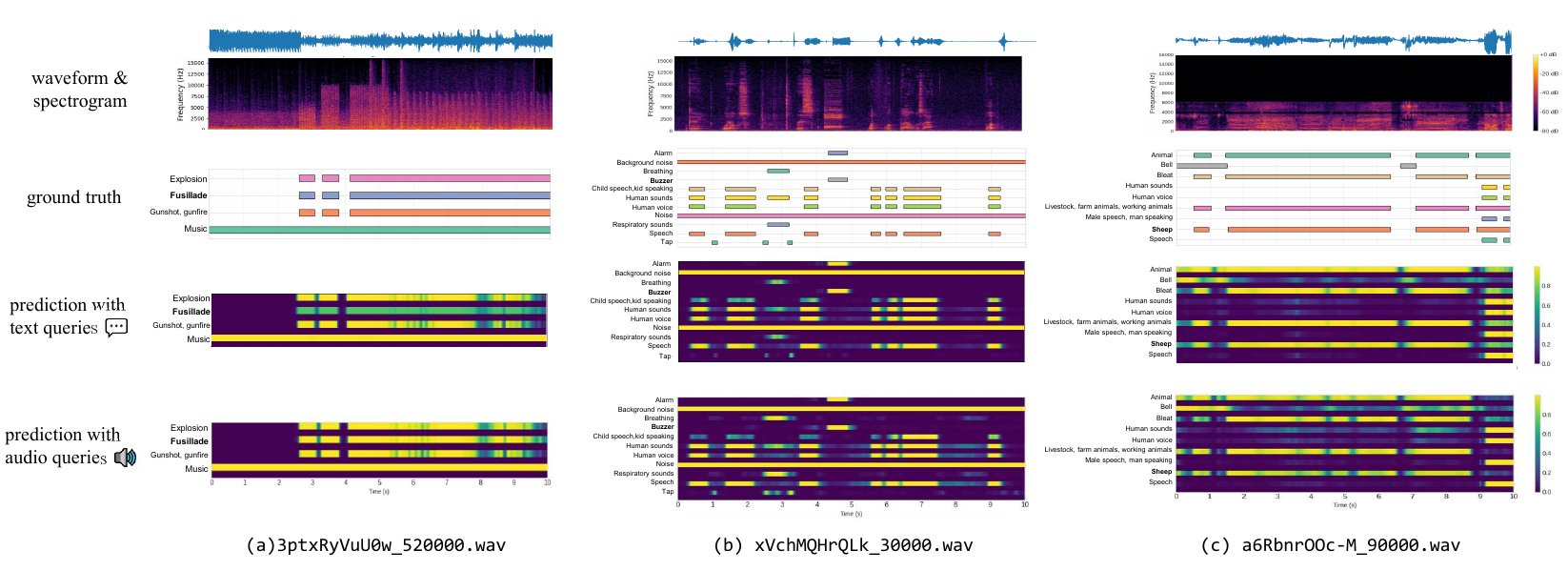}
  \caption{Qualitative results of DASM. The model is trained under the AS-partial setting with an HTS-AT backbone and multi-modal queries.
  The audio samples are randomly selected from the AudioSet evaluation set, and those in bold are novel classes.
}
  \label{fig:demo}
\end{figure*}

\subsection{Ablation Study}
In this section, we conduct ablation experiments to assess the contribution of different components in DASM.  

\paragraph{\textbf{Ablation of Audio Query Duration}}
To investigate how audio query length affects performance, we evaluate with different duration audio queries.
Specifically, we evaluate DASM under the AS-partial setting on 12 novel classes with original query duration ranges from 5–6 minutes, then progressively reduce the query duration through random sampling.
As shown in Fig.~\ref{fig:duration}, just a few minutes of audio is sufficient for DASM to construct effective audio queries. This is valuable for scenarios with limited audio resources.

\paragraph{\textbf{Ablation of the Decoder}}  
The dual-stream decoder is the core component of DASM, where the cross-modality event decoder fuses query vectors with audio features to produce clip-level predictions, while the context network models the temporal dependencies to generate frame-level predictions. 
The results of the ablation study, using the HTS-AT backbone and text queries, are presented in Tab.~\ref{tab:ablation_text_query}. 
Comparing Model 1 with Model 0, we observe a substantial drop in performance on open-vocabulary tasks when the event decoder is removed, reducing the model’s generalization capability to a level similar to CLAP-based architectures. 
This highlights the critical role of the event decoder in facilitating cross-modal feature fusion.  
In Model 2, we remove the Conformer based context network, and observe a notable performance degradation in both closed-set and open-vocabulary tasks. 
This result underscores the importance of the context network in modeling the temporal relationships within the audio feature sequence, which is crucial for event localization.
Models 3 and 4 examine the impact of clip-level predictions.
Removing the clip-level loss from Eq.~\ref{equ:loss} or the clip-level prior from Eq.~\ref{eq:y_frame_final} leads to a significant performance drop across all evaluation settings. This demonstrates the importance of both the clip-level prior and the clip-level loss, further validating the effectiveness of the factorized computation of the decoder.

\paragraph{\textbf{Ablation of Inference-Time Masking}}  
To analyze the impact of base-class visibility in open-vocabulary tasks, we investigate two  self-attention masking strategies in the event decoder during inference.
The results, shown in Tab.~\ref{tab:inference}, report PSDS$_r$ scores under the AS-partial setting.
When base classes are invisible to novel classes as illustrated in Fig~\ref{fig:mask}~(a), the model almost loses its ability to detect novel classes. 
This suggests that DASM heavily relies on base class queries for novel class detection. 
One possible explanation is that maintaining base class visibility allows the model to exploit  semantic relationships and hierarchical structures between novel and base classes, as further illustrated in the next section. 
Another possible interpretation is that base class queries serve as a set of basis vectors in the query space, enabling generalization to novel classes through their combinations, and this generalization capability is lost when base classes are masked.

\subsection{Qualitative Results}
We visualize DASM’s detection results of several examples in Fig.~\ref{fig:demo}.  
Overall, DASM exhibits strong capability in detecting sound events, regardless of using text or audio queries. 
The results for provided examples demonstrate DASM's robustness in complex acoustic environments, where multiple overlapping events with varying duration and characteristics coexist—some of which are novel and have never been encountered during training.
Notably, multiple labels can be attached to the same sound instance, due to label augmentation with AudioSet Ontology and the inherent semantic similarity between labels.
Our model effectively captures these hierarchical and relational structures in the semantic space. 
For instance, in Fig.~\ref{fig:demo}~(a), we observe that the predictions for the novel class ``Fusillade'' closely align with those of the base classes ``Explosion'' and ``Gunshot, gunfire'', both of which can be considered hypernyms of ``Fusillade''.  

\section{Conclusion}
In this paper, we propose DASM, a query-based framework for open-vocabulary sound event detection that enables flexible detection of arbitrary sound events guided by either text or audio queries.
DASM formulates sound event detection as a frame-level retrieval task and integrates three key components: a CLAP-based query generation module supporting multi-modal queries, an audio encoder for fine-grained audio representation, and a dual-stream decoder that decouples event recognition and localization through a cross-modality event decoder and a context network.
Extensive experiments on AudioSet-Strong and DESED demonstrate that DASM achieves strong generalization to novel classes while maintaining high localization accuracy comparable to closed-set SED models.  
Looking forward, we would like to extend the framework to support additional modalities, such as image and video.
In addition, we hope to combine our framework with multimodal large language models, to further enhance its ability to understand complex textual descriptions.

\FloatBarrier
\newpage

\bibliographystyle{ACM-Reference-Format}
\balance
\bibliography{sample-base}

\end{document}